\documentclass[a4paper,11pt]{article}
\pdfoutput=1 

\usepackage{jinstpub} 

\usepackage{lineno}
\usepackage{xcolor}
\usepackage{ulem}

\title{\boldmath Evolution of the electrical characteristics of the ATLAS18 ITk strip sensors with HL-LHC radiation exposure range}

\author[a,b,1]{J. Fern\'andez-Tejero,\note{Corresponding author.}}
\author[c]{\`E.~Bach,}
\author[d]{V.~Cindro,}
\author[e]{V.~Fadeyev,}
\author[f]{P.~Federi\v{c}ov\'a,}
\author[c]{C.~Fleta,}
\author[g]{S.~Hirose,}
\author[f]{J.~Kroll,}
\author[d]{I.~Mandi\'c,}
\author[g]{K.~Maeyama,}
\author[f]{M.~Mike\v{s}t\'ikov\'a,}
\author[a,b]{L.~Poley,}
\author[a,b]{B.~Stelzer,}
\author[f]{P.~Tuma,}
\author[c]{M.~Ull\'an,}
\author[h]{Y.~Unno}

\affiliation[a]{Department of Physics, Simon Fraser University, 8888 University Drive, Burnaby, B.C. V5A 1S6, Canada}
\affiliation[b]{TRIUMF, 4004 Wesbrook Mall, Vancouver, B.C. V6T 2A3, Canada}
\affiliation[c]{Centro Nacional de Microelectr\'onica (IMB-CNM, CSIC), Campus UAB-Bellaterra, 08193 Barcelona, Spain}
\affiliation[d]{Experimental Particle Physics Department, Jo{\v z}ef Stefan Institute, Jamova 39, SI-1000 Ljubljana, Slovenia}
\affiliation[e]{Santa Cruz Institute for Particle Physics (SCIPP), University of California, Santa Cruz, CA 95064, USA}
\affiliation[f]{Academy of Sciences of the Czech Republic, Institute of Physics, Na Slovance 2, 18221 Prague 8, Czech Republic}
\affiliation[g]{Institute of Pure and Applied Sciences, University of Tsukuba, 1-1-1 Tennodai, Tsukuba, Ibaraki 305-8571, Japan}
\affiliation[h]{Institute of Particle and Nuclear Study, High Energy Accelerator Research Organization (KEK), 1-1 Oho, Tsukuba, Ibaraki 305-0801, Japan}

\emailAdd{Xavi.Fdez@cern.ch}

\abstract{The objective of the study is to evaluate the evolution of the performance of the new ATLAS Inner-Tracker (ITk) strip sensors as a function of radiation exposure, to ensure the proper operation of the upgraded detector during the lifetime of the High-Luminosity Large Hadron Collider (HL-LHC). Full-size ATLAS ITk Barrel Short-Strip (SS) sensors with final layout design, ATLAS18SS, have been irradiated with neutrons and gammas, to confirm the results obtained with prototypes and miniature sensors during the development phase. The irradiations cover a wide range of fluences and doses that ITk will experience, going from 1x10$^{13}$ n$_{eq}$/cm$^{2}$ and 0.49 Mrad, to 1.6x10$^{15}$ n$_{eq}$/cm$^{2}$ and 80 Mrad. The split irradiation enables a proper combination of fluence and dose values of the HL-LHC, including a 1.5 safety factor. A complete electrical characterization of the key sensor parameters before and after irradiation is presented, studying the leakage current, bulk capacitance, single-strip and inter-strip characteristics. The results confirm the fulfilment of the ATLAS specifications throughout the whole experiment. The study of a wide range of fluences and doses also allows to obtain detailed results, such as the frequency dependence of the bulk capacitance measurements for highly irradiated sensors, or the evolution of the punch-through protection and inter-strip resistance with radiation.
}

\keywords{Radiation-hard strip sensors, Si microstrip and pad detectors, Particle tracking detectors (Solid-state detectors)}

\begin{document}
\maketitle
\flushbottom


\section{ATLAS ITk strip sensors and specifications}

In 2021, the ATLAS collaboration started the production phase of the new Inner-Tracker (ITk) strip sensors~\cite{UnnoPre-production}, so-called ATLAS18, that should be able to withstand the extreme radiation conditions expected for the forthcoming High-Luminosity Large Hadron Collider (HL-LHC)~\cite{HLLHC}. The new all-silicon ITk detector~\cite{ATLASTDR-strips} will reach unprecedented accumulated fluences and ionizing doses, caused by the increase of the total integrated luminosity. Previous studies with prototypes and miniature sensors~\cite{XaviThesis, XaviIFX, MarcelaA17} showed that these severe radiation conditions can modify some of the electrical characteristics of the strip sensors in working conditions.  

\begin{figure}
  \centering
  \includegraphics[width=0.5\textwidth]{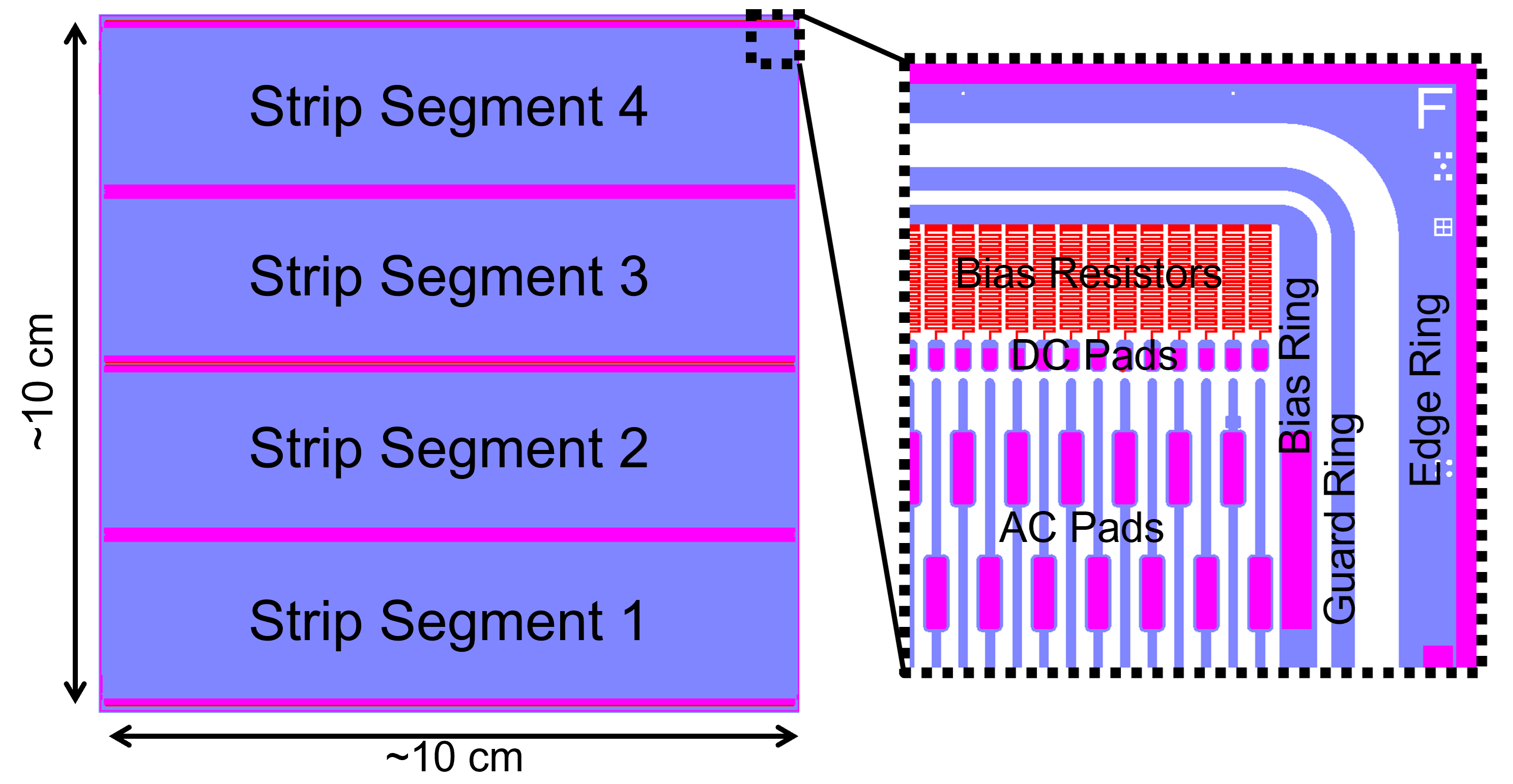}
{\caption{ATLAS ITk Short-Strip sensor, so-called ATLAS18SS, showing dimensions and main structures. \label{fig:Chapter0:sensor}}}
\end{figure}

The strip region of the new ATLAS ITk consists of a four-layer Barrel section, and one End-cap section on each side, composed of six disks per side. The Barrel section contains 2 types of square-shaped sensors with different strip length, i.e. Short-Strip (SS) and Long-Strip (LS), while the End-cap section is composed of 6 sensor types with the strips oriented radially to the beam axis, i.e. Ring 0 (R0) to Ring 5 (R5). ITk strip sensors are fabricated by Hamamatsu Photonics KK, using Float Zone p-type 6-inch silicon wafers, with 320 $\mu$m thickness. Full-size ATLAS18 Barrel SS sensors, with dimensions of 97621 x 97950 $\mu$m$^{2}$, have been used for this study. The active area is composed of 4 segments with 1280 parallel strips per segment, with a strip pitch of 75.5 $\mu$m and length of 2.4 cm, isolated by p-stop structures. Strips have AC and DC pads to contact the strip metal and strip implant, respectively, and are connected through a poly-silicon bias resistor to the bias ring surrounding the active area, followed by a guard ring (n+) and an edge ring (p+), responsible of shaping the electric field avoiding the appearance of an inversion region at the edge of the device (Figure~\ref{fig:Chapter0:sensor}). Each strip is also equipped with a Punch-Through Protection (PTP) structure, where the distance between the strip implant and the bias ring implant is reduced, shorting the implant to the grounded bias ring when a certain voltage threshold is reached, protecting the coupling capacitor.

With the aim to evaluate the performance of the new ITk strip sensors, and confirm the results obtained during the development phase with miniature sensors, full-size ATLAS18SS sensors have been irradiated with neutrons and gammas. The irradiations cover a wide range of fluences and doses, from equivalent to the ones expected from the first days of HL-LHC operation, to the end of the lifetime of the experiment~\cite{ATLASTDR-strips}. This study presents a complete characterization of the key electrical characteristics of the sensors before and after irradiation. The Technical Specifications for the ATLAS ITk strip sensors~\cite{UnnoPre-production}, defining the thresholds established by the collaboration for the electrical parameters evaluated, are summarized in Table~\ref{tab:specs}, providing performance benchmarks.

\begin{table}
\caption{\label{tab:specs} Electrical specifications for ATLAS ITk strip sensors (ATLAS18), before irradiation and at target fluence (1.6x10$^{15}$ n$_{eq}$/cm$^{2}$), which includes a 1.5 safety factor.}
\smallskip
\scalebox{0.9}{
\begin{tabular}{|c|c|c|}
\hline
\textbf{Electrical}&\textbf{Before}&\textbf{At target fluence}\\
\textbf{parameter}&\textbf{irradiation}&\textbf{(1.6x10$^{15}$ n$_{eq}$/cm$^{2}$)}\\
\hline
Breakdown voltage (V$_{bd}$) & $>$500 V & $>$500 V \\
\hline
Leakage current at 500 V (I$_{500}$) & $<$0.1 $\mu$A/cm$^{2}$ (20$^\circ$C) & $<$100 $\mu$A/cm$^{2}$ (-20$^\circ$C) \\
\hline
Full depletion voltage (V$_{fd}$) & $<$350 V & - \\
\hline
Coupling capacitance (C$_{coupl}$) & $\geq$20 pF/cm & - \\
\hline
Bias resistance (R$_{bias}$) & 1.5$\pm$0.5 M$\Omega$ & 1.8$\pm$0.5 M$\Omega$ \\
\hline
Punch-through voltage (V$_{PT}$) & - & - \\
\hline
Inter-strip capacitance (C$_{int}$) & $<$1 pF/cm (-300 V) & $<$1 pF/cm (-300 V) \\
\hline
Inter-strip resistance (R$_{int}$) & $>$10xR$_{bias}$ (-400 V) & $>$10xR$_{bias}$ (-500 V) \\
\hline
\end{tabular}
}
\end{table}

\section{Neutron and gamma irradiations}\label{sec:irradiations}

The objective of the irradiation plan is to cover the HL-LHC radiation exposure range expected from the very first days of the HL-LHC operation, to the target at the end of the lifetime of the experiment, including a 1.5 safety factor. Sensors were irradiated with neutrons plus the expected equivalent dose with gammas, enabling a proper combination of fluence and dose values, which is not possible with a proton irradiation in 10 to 100 MeV kinetic energies.

In total, 5 full-size ATLAS18SS sensors were irradiated with neutrons at the TRIGA-Mark-III nuclear reactor of the Jo\v{z}ef Stefan Institute (JSI) in Ljubljana (Slovenia), and with $^{60}$Co gammas at UJP Praha by the Institute of Physics of the Czech Academy of Science (FZU) in Prague (Czech Republic). Irradiations were performed in two rounds (Table~\ref{tab:irrad}) with the aim to compare the influence of neutron and neutron+gamma irradiations. A first neutron irradiation of 3 sensors to the 3 lowest fluences, i.e. 1x10$^{13}$, 5x10$^{13}$ and 1x10$^{14}$ n$_{eq}$/cm$^{2}$, and a second round of neutron and gamma irradiations, adding 2 more fluences, i.e. 5.1x10$^{14}$ and 1.6x10$^{15}$ n$_{eq}$/cm$^{2}$, and the corresponding gamma doses for all the sensors irradiated, i.e. 0.49, 2.4, 4.9, 24.8 and 80 Mrad.

Additionally, a non-irradiated full-size ATLAS18SS sensor was used to evaluate the leakage current and bulk capacitance before irradiation, while a miniature sensor composed by 104 parallel strips with pitch and length identical to the ATLAS18SS sensors, so-called MiniSS\footnote{There are three types of miniature sensors in the ATLAS18 wafers, Mini (0.8 cm), MiniSS (2.4 cm), and MiniLS (4.8 cm), with the strip length in the parenthesis.}, was used to evaluate the single strip and inter-strip characteristics.

\begin{table}
\caption{\label{tab:irrad} Devices under test and irradiations.}
\smallskip
\scalebox{0.9}{
\begin{tabular}{|c|c|c|}
\hline
\textbf{Sensor Type}&\textbf{1st Irradiation}&\textbf{2nd Irradiation}\\
&\textbf{(n: neutrons)}&\textbf{(n: neutrons, g: gammas)}\\
\hline
ATLAS18SS & pre-irrad tests & pre-irrad tests \\
\hline
ATLAS18-MiniSS & pre-irrad tests & pre-irrad tests \\
\hline
ATLAS18SS & 1x10$^{13}$ n$_{eq}$/cm$^{2}$ (n) & 0.49 Mrad (g) \\
\hline
ATLAS18SS & 5x10$^{13}$ n$_{eq}$/cm$^{2}$ (n) & 2.4 Mrad (g) \\
\hline
ATLAS18SS & 1x10$^{14}$ n$_{eq}$/cm$^{2}$ (n) & 80 Mrad (g) \\
\hline
ATLAS18SS & - & 5.1x10$^{14}$ n$_{eq}$/cm$^{2}$ (n) + 24.8 Mrad (g) \\
\hline
ATLAS18SS & - & 1.6x10$^{15}$ n$_{eq}$/cm$^{2}$ (n) + 80 Mrad (g) \\
\hline
\end{tabular}
}
\end{table}

\section{Test methods}\label{sec:tests}

\begin{figure}
  \centering
  \includegraphics[width=0.6\textwidth]{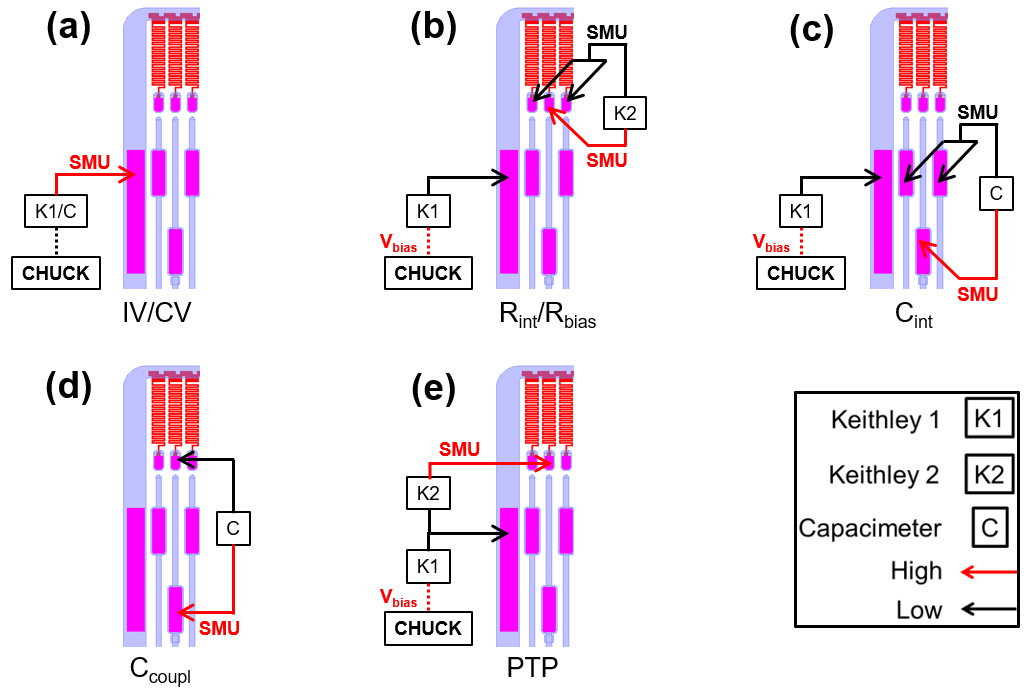}
{\caption{Test methods to evaluate the sensor leakage current (IV) and bulk capacitance (CV) (a), inter-strip (R$_{int}$) and bias (R$_{bias}$) resistance (b), inter-strip capacitance (C$_{int}$) (c), coupling capacitance (C$_{coupl}$) (d) and punch-through protection (PTP) (e). Figure adapted from~\cite{XaviThesis}. \label{fig:Chapter0:test-methods}}}
\end{figure}

The test methods used for this study are based on the Technical Specifications for the ATLAS ITk strip sensors~\cite{UnnoPre-production}, with schematic representations in Figure~\ref{fig:Chapter0:test-methods}. All the electrical tests were performed in a shielded probe station in a dry environment (RH$<$5\%), using a nitrogen flow. Measurements before irradiation were performed at 20$^\circ$C and after irradiation at -20$^\circ$C, using a thermal chuck for the device cooling. Irradiated sensors were annealed before the characterization for 80 minutes at 60$^\circ$C.

The leakage current and bulk capacitance were measured up to -500 V, i.e. 500 V reverse bias voltage relative from the bias ring to the backplane, in steps of 10 V, using for the capacitance measurements a CR-series configuration and different frequencies. The coupling capacitance was measured with the sensors unbiased, CR-parallel configuration and at 1 kHz. While the effectiveness of the PTP structure was evaluated applying a voltage sweep (V$_{test}$) up to 30 V to the DC pad (strip implant) and measuring the induced current (I$_{test}$) between the strip implant and the bias ring implant, with the sensor biased to -500 V. The effective resistance (R$_{eff}$) was then calculated from the equivalent circuit composed by the bias resistance (R$_{bias}$) in parallel with the punch-through resistance (R$_{PT}$)~\cite{PTP}:

\[R_{eff}=\frac{V_{test}}{I_{test}}=\bigg(\frac{1}{R_{bias}}+\frac{1}{R_{PT}}\bigg)^{-1}\]

Then, the punch-through voltage (V$_{PT}$) can be extracted for the condition \textit{R$_{PT}$ = R$_{bias}$}, i.e. \textit{R$_{eff}$ = R$_{bias}$/2}.

For the inter-strip resistance and capacitance, the measurements were performed considering the first neighbouring strips at both sides of the strip under test, using the DC pads (contacting the strip implant) for the resistance measurement, and the AC pads (contacting the strip metal) for the capacitance. For the resistance measurement, a voltage sweep between -1 and 1 V was applied to the strip under test, measuring the current induced in the neighbouring strips to extract the inter-strip resistance, with the sensor biased to -400 V before irradiation, and -500 V after irradiation. While for the capacitance measurement, an LCR meter was used in CR-parallel configuration, with a test frequency of 100 kHz, and the sensor biased to -300 V. On the other hand, the inter-strip resistance set-up was also used to measure the bias resistance, but in this case measuring the current induced in the strip under test, and extracting the resistance of the poly-silicon bias resistor.

\section{Results}

\subsection{Leakage current and bulk capacitance}

\begin{figure}
  \centering
  \includegraphics[width=0.4\textwidth]{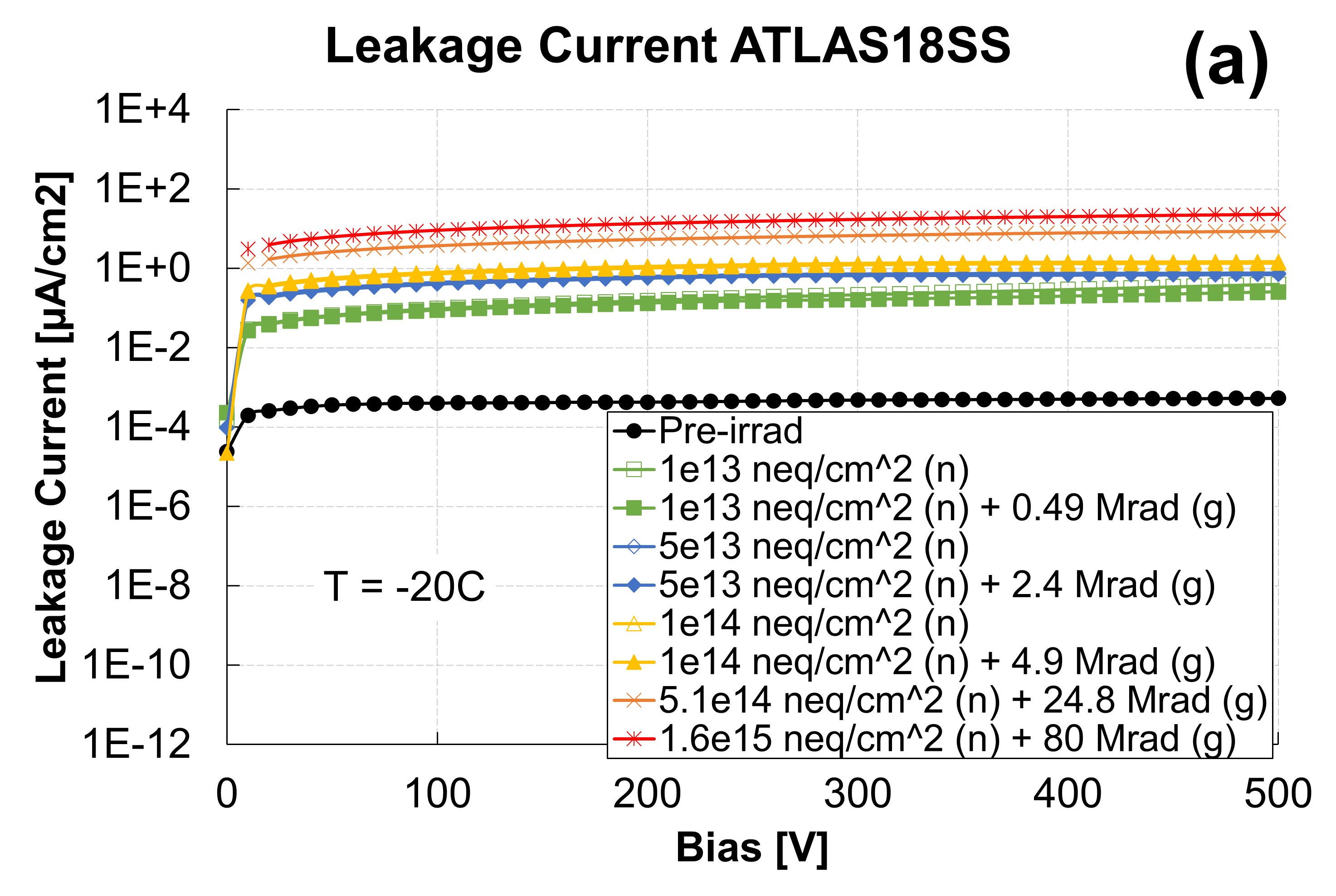}
  \includegraphics[width=0.4\textwidth]{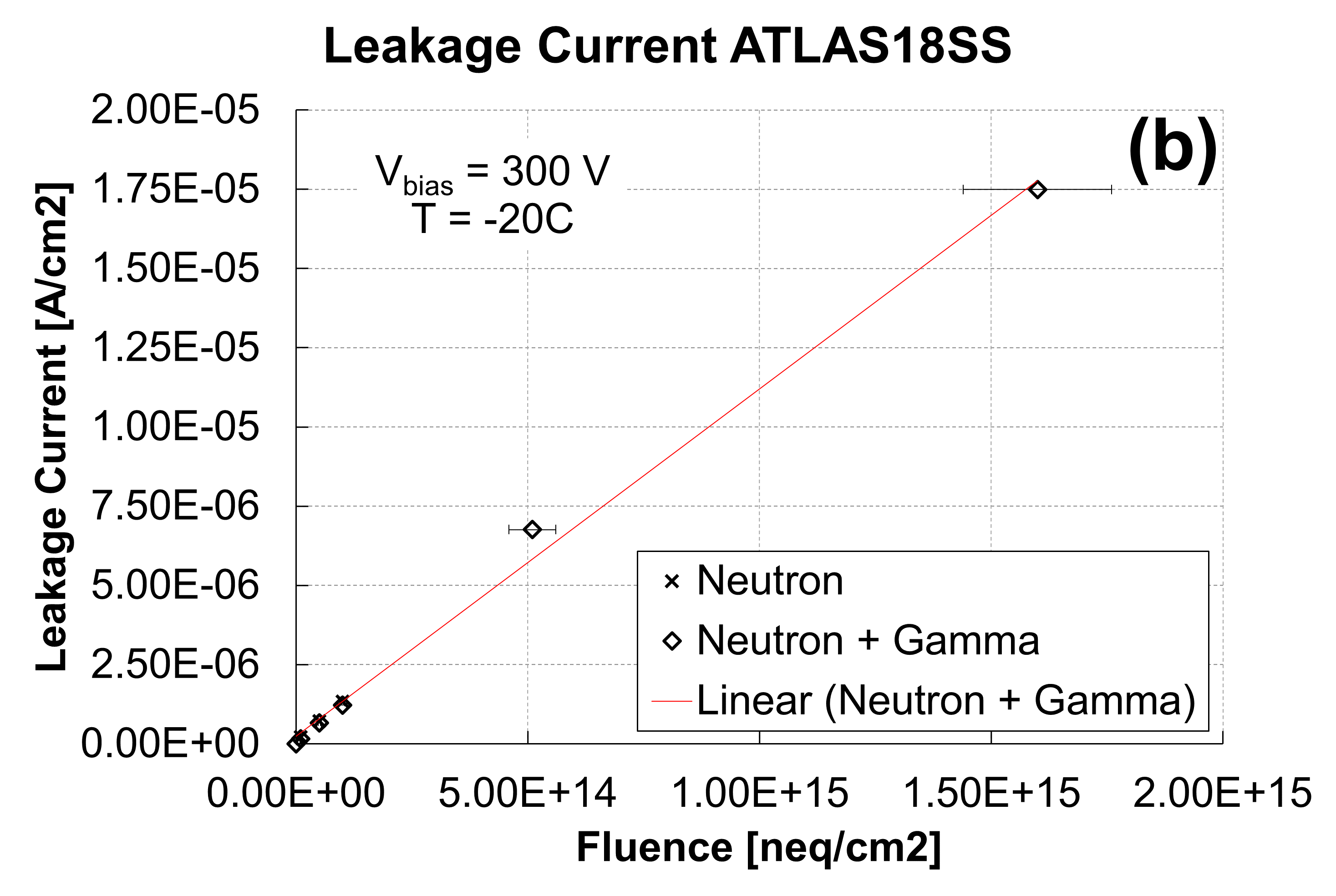}
{\caption{Reverse leakage current of neutron, and neutron+gamma, irradiated ATLAS18SS sensors as a function of bias voltage (a), and fluence (b). \label{fig:Chapter1:leakage_current}}}
\end{figure}

No breakdown is observed below -500 V before or after irradiation (Figure~\ref{fig:Chapter1:leakage_current}), showing a linear increase of leakage current as a function of fluence, corresponding to a current related damage rate ($\alpha$) of 3.34x10$^{-17}$ A/cm, consistent with the damage rates of 4-5x10$^{-17}$ A/cm expected for fluences below 1x10$^{16}$ n$_{eq}$/cm$^{2}$~\cite{damage-rate}. The leakage current remains below $<$0.1 $\mu$A/cm$^{2}$ before irradiation, and below $<$100 $\mu$A/cm$^{2}$ at target fluence, also in agreement with the ATLAS ITk specifications (Table~\ref{tab:specs}). On the other hand, no clear difference in leakage current was observed for sensors irradiated only with neutrons compared with neutrons+gammas.

\begin{figure}
  \centering
  \includegraphics[width=0.3\textwidth]{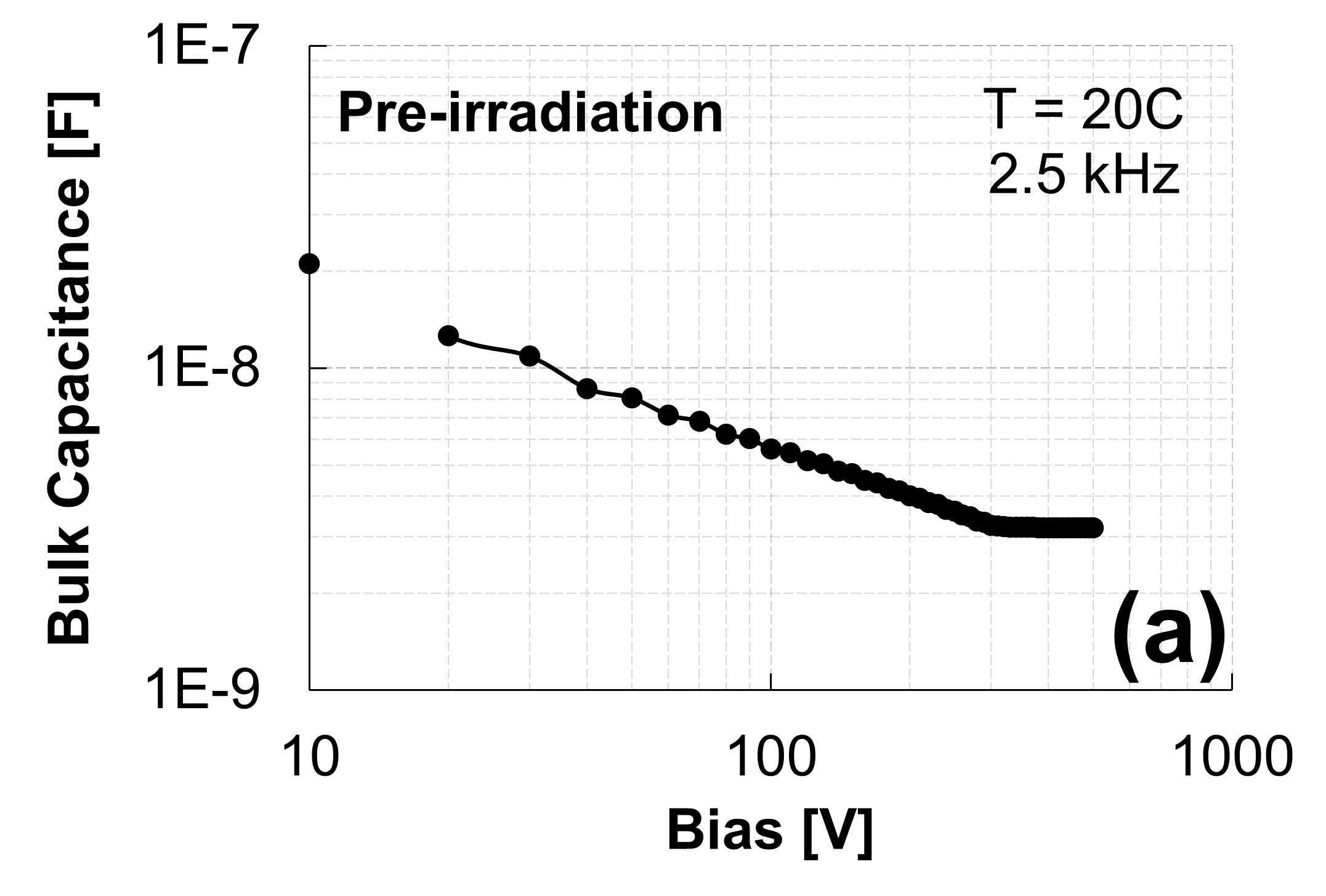}
  \includegraphics[width=0.3\textwidth]{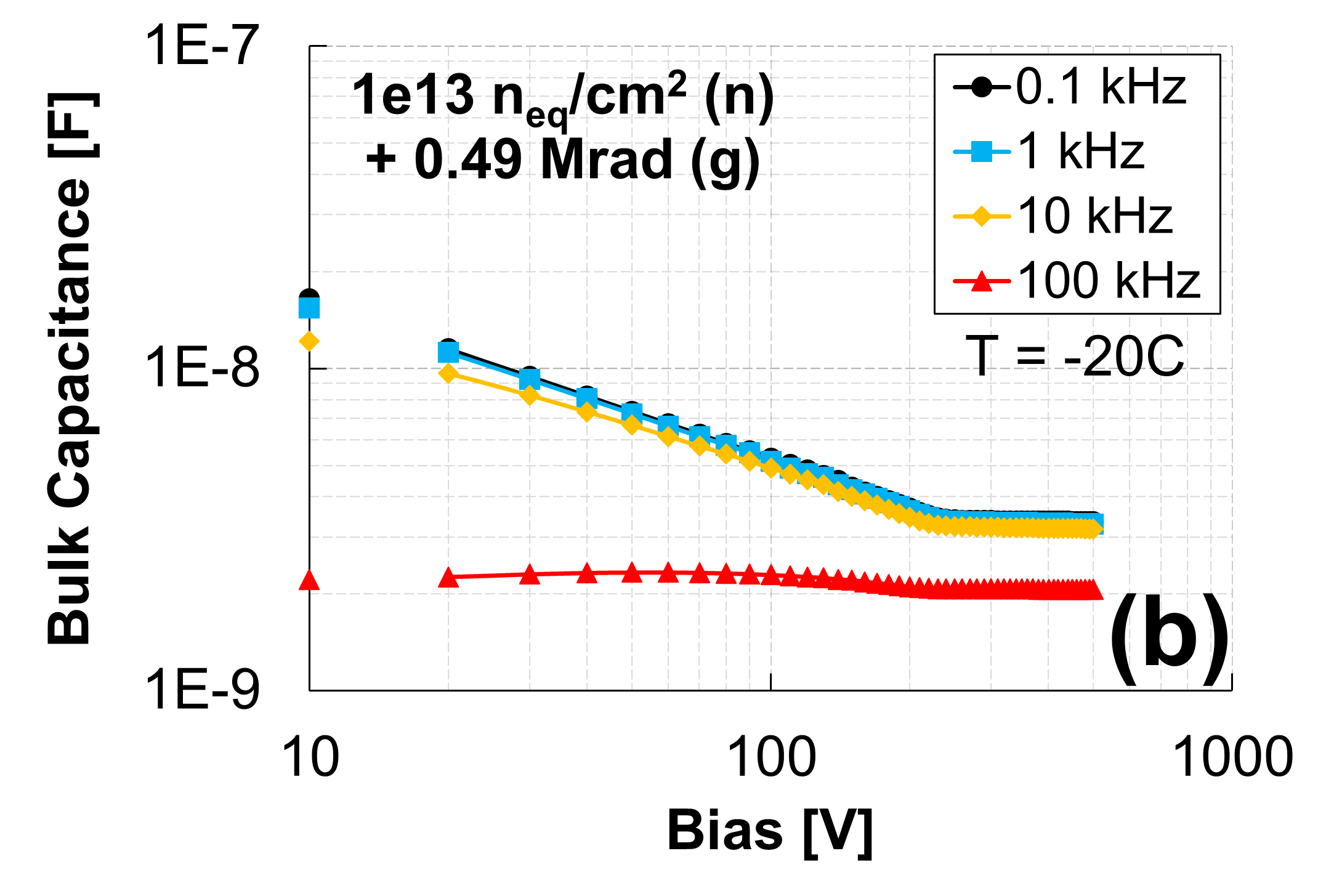}
  \includegraphics[width=0.3\textwidth]{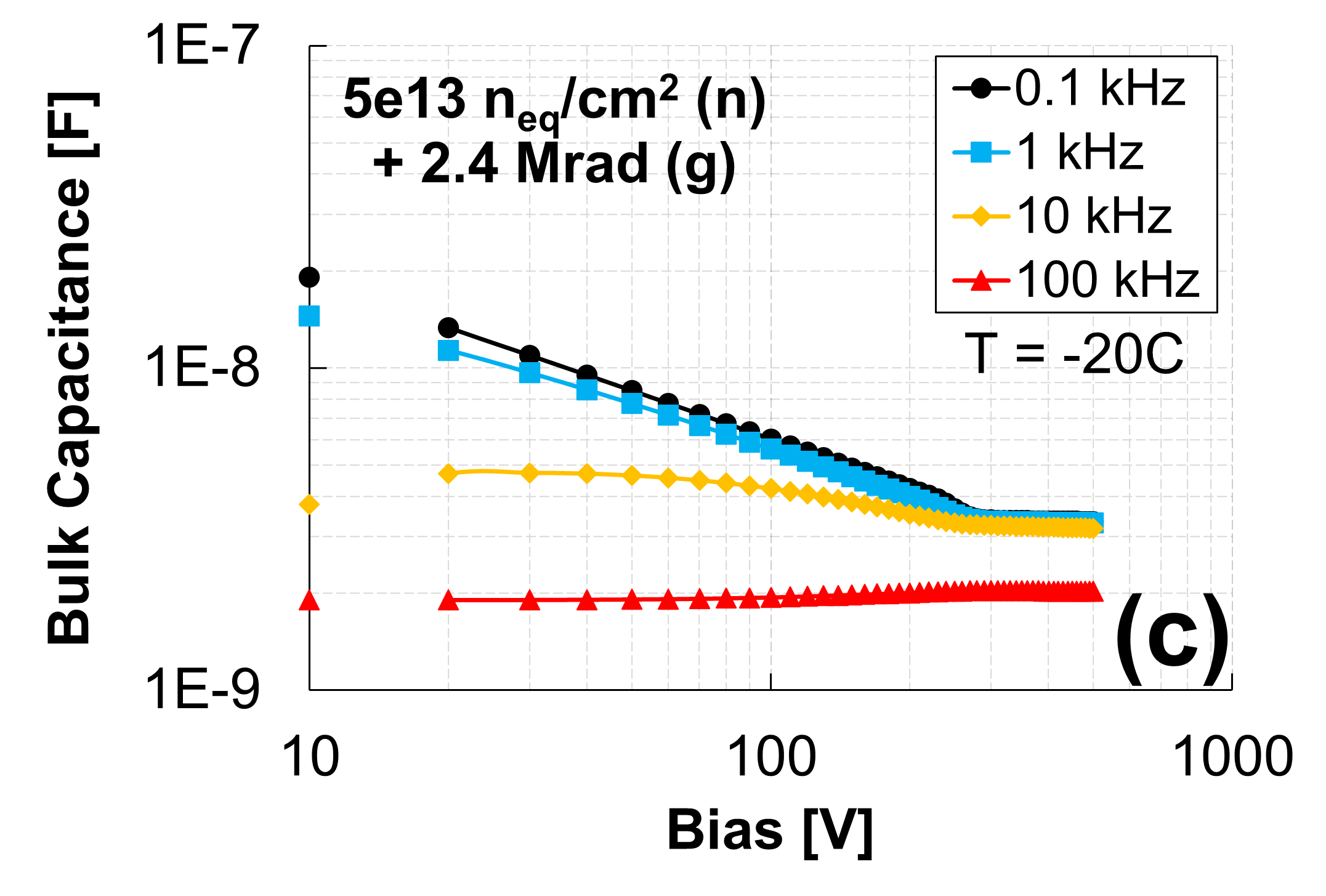}
  \includegraphics[width=0.3\textwidth]{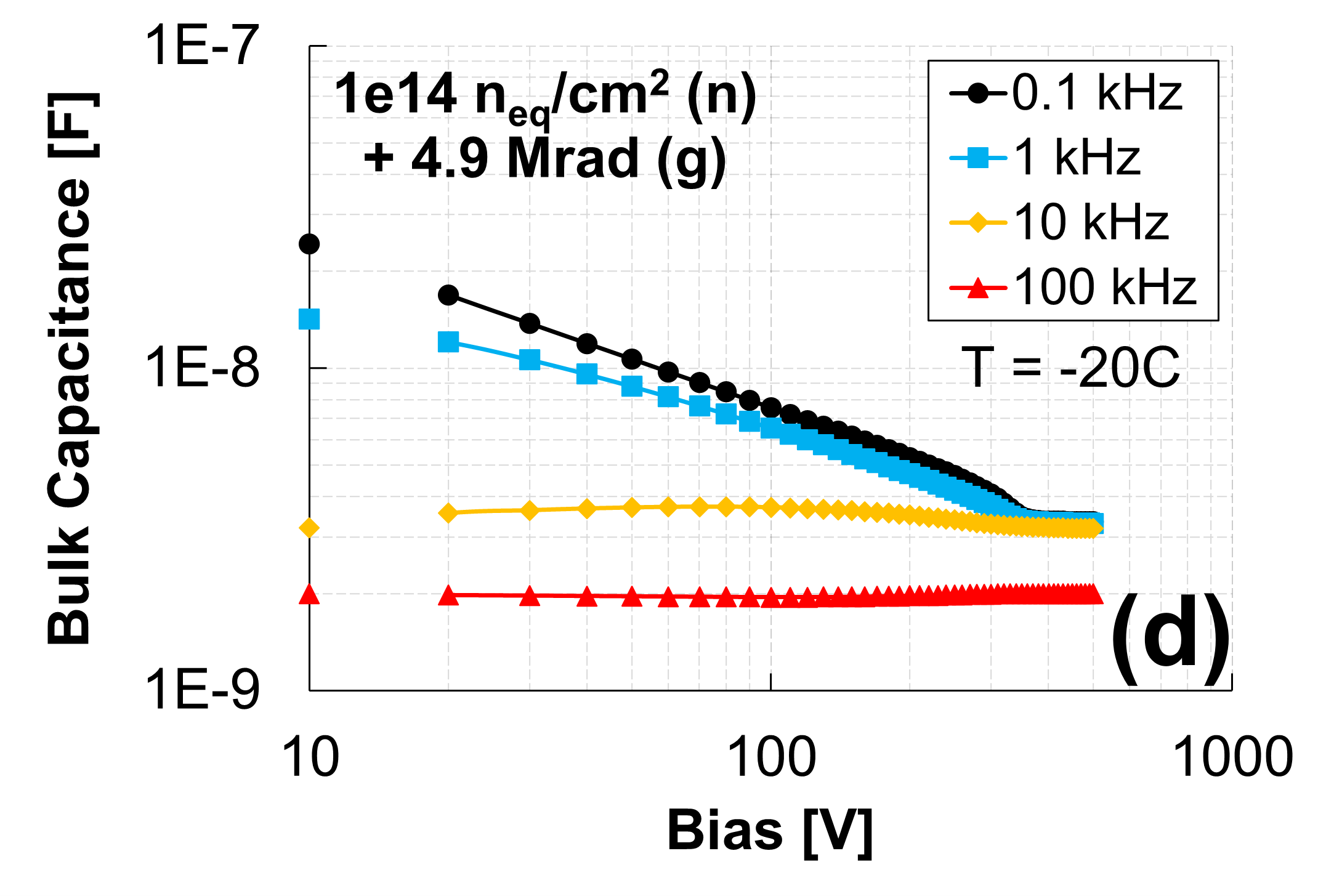}
  \includegraphics[width=0.3\textwidth]{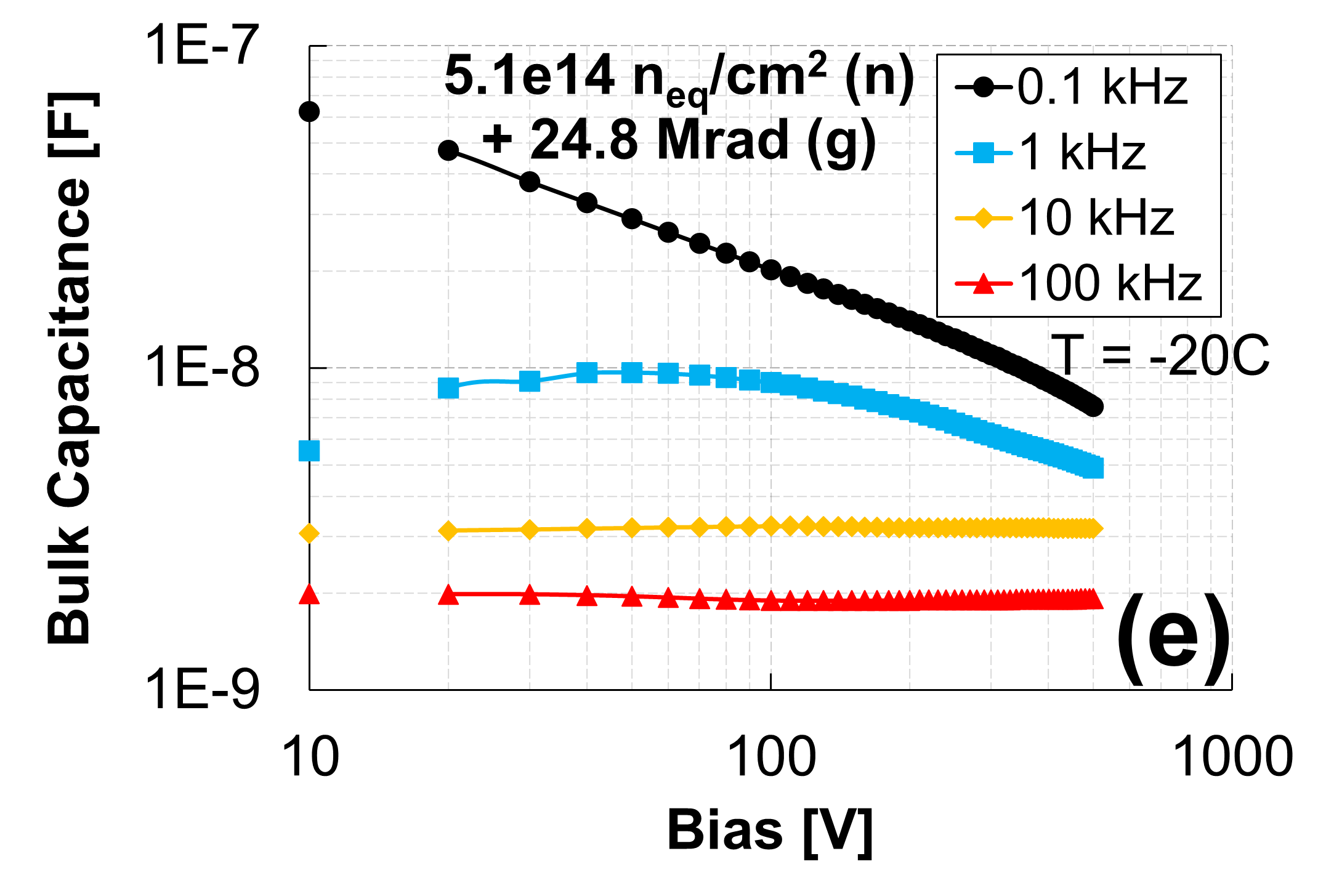}
  \includegraphics[width=0.3\textwidth]{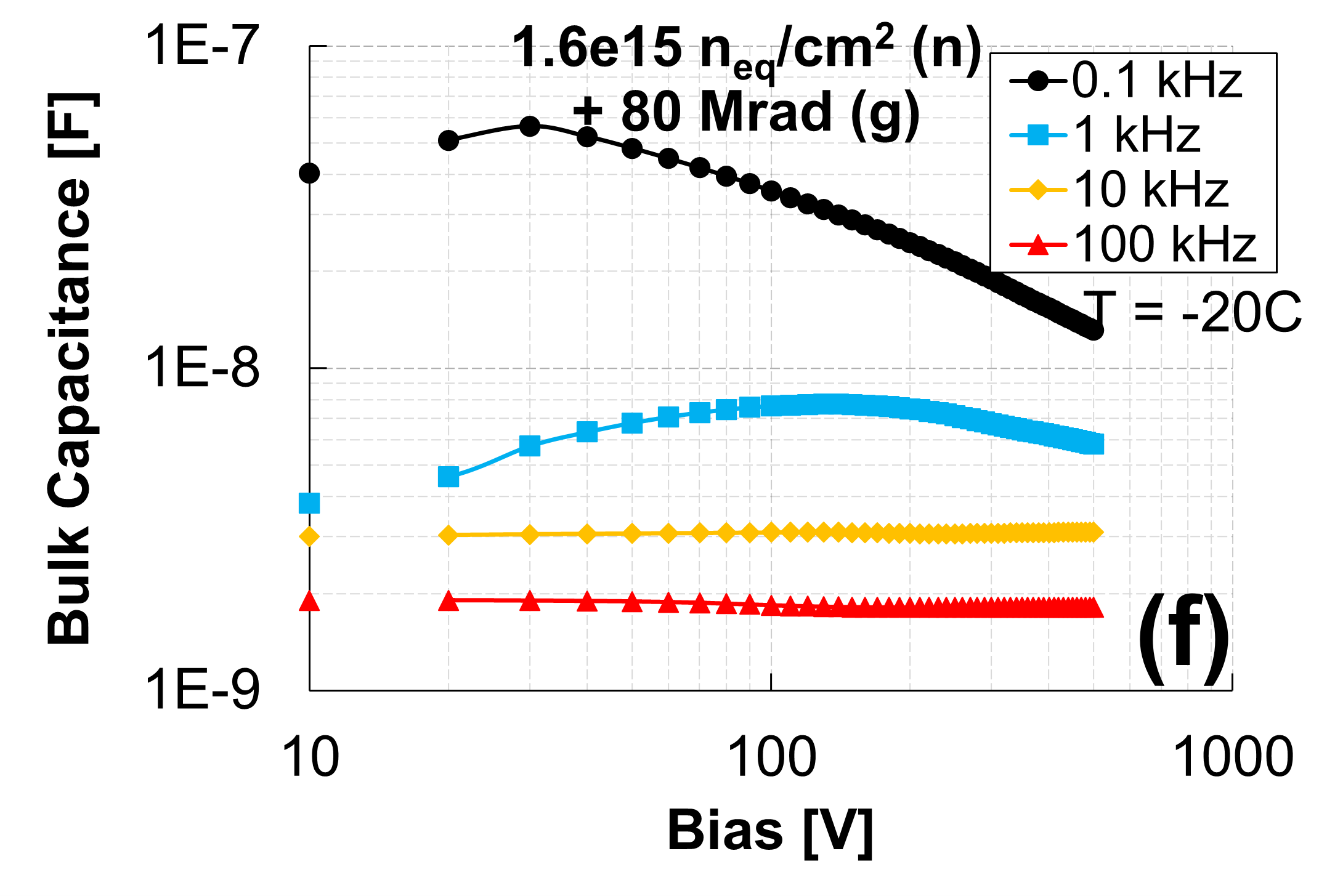}
  \includegraphics[width=0.3\textwidth]{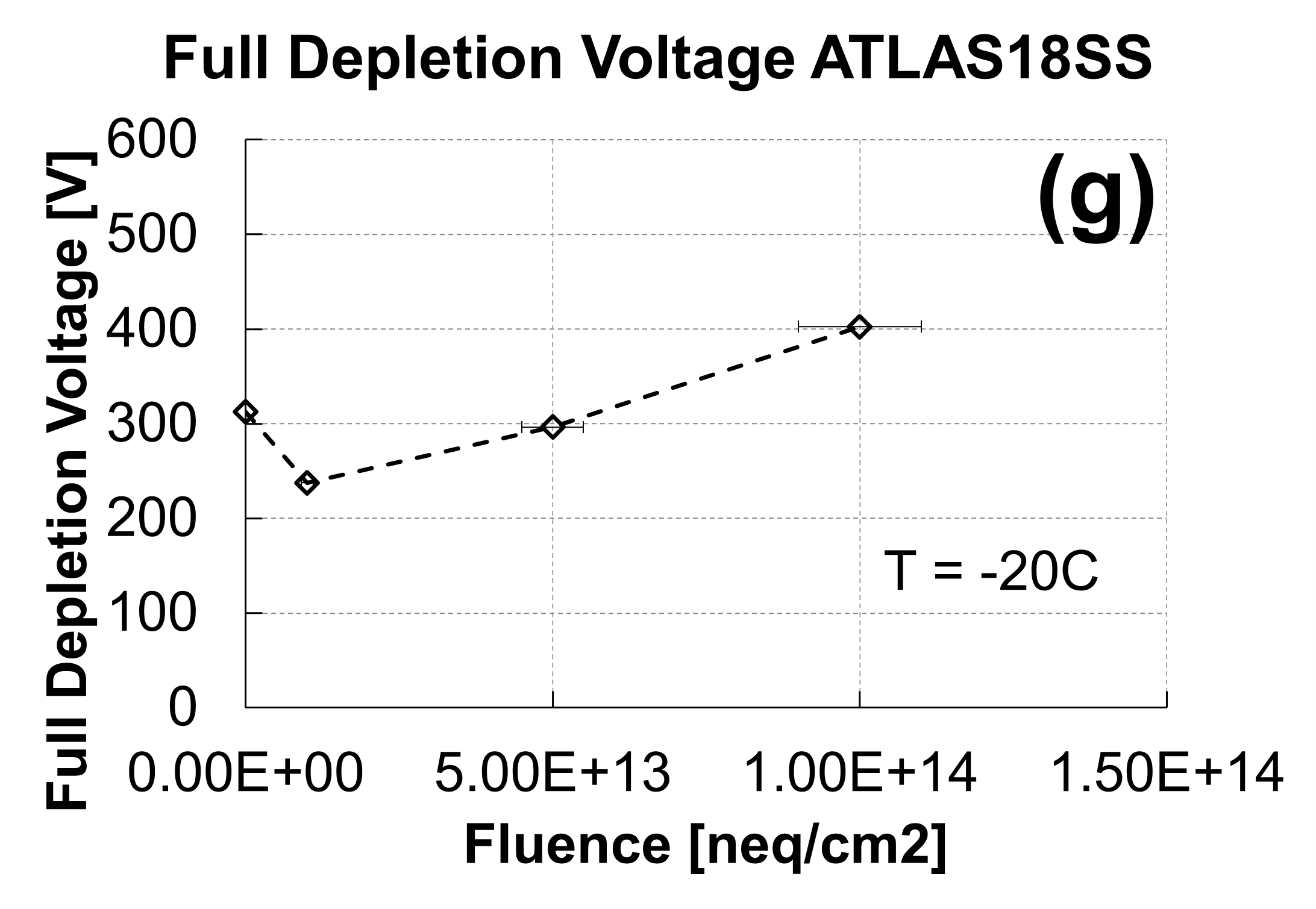}
{\caption{Bulk capacitance at different frequencies of ATLAS18SS sensors irradiated to different neutron+gamma fluences (a-f), and full depletion voltage as a function of fluence up to 1x10$^{14}$ n$_{eq}$/cm$^{2}$ (g). \label{fig:Chapter1:bulk_capacitance}}}
\end{figure}

The bulk capacitance before irradiation was measured at 2.5 kHz frequency, showing a full depletion voltage at 312 V (Figure~\ref{fig:Chapter1:bulk_capacitance}(a)), in agreement with the specifications ($<$350 V). For the irradiated sensors, a more extensive study was performed to investigate the evolution of the frequency dependence after irradiation. Figures~\ref{fig:Chapter1:bulk_capacitance}(b-f) show the bulk capacitance obtained for the different neutron+gamma irradiation fluences when measured at frequencies of 0.1, 1, 10 and 100 kHz. The results show that lower frequencies are needed to extract the full depletion voltage when measuring highly irradiated sensors. As suggested in previous studies with lower fluences up to 1x10$^{14}$ n$_{eq}$/cm$^{2}$~\cite{frequencyCV}, frequency should be low enough to interact with the deep traps created during neutron irradiation. Interestingly, measurements at frequencies below 10 kHz show lower full depletion voltages below 5x10$^{13}$ n$_{eq}$/cm$^{2}$ fluences, not studied previously with prototypes, followed by an increase for higher fluences (Figure~\ref{fig:Chapter1:bulk_capacitance}(g)). Further investigations are needed to understand the reduction of full depletion voltage for low fluences, which could be attributed to the "acceptor removal" effect~\cite{acceptor-removal}. Additionally, the measurements show that sensors irradiated above 5.1x10$^{14}$ n$_{eq}$/cm$^{2}$ require bias voltages above -500 V to fully deplete silicon bulk. The capacitance measured at high frequency remains flat with voltage. This implies that the under-depletion will not impact the readout noise in the tracker, where the front-end circuitry includes a high-frequency bandpass~\cite{frontend-frequency}.

\subsection{Inter-strip characterization}

\begin{figure}
  \centering
  \includegraphics[width=0.4\textwidth]{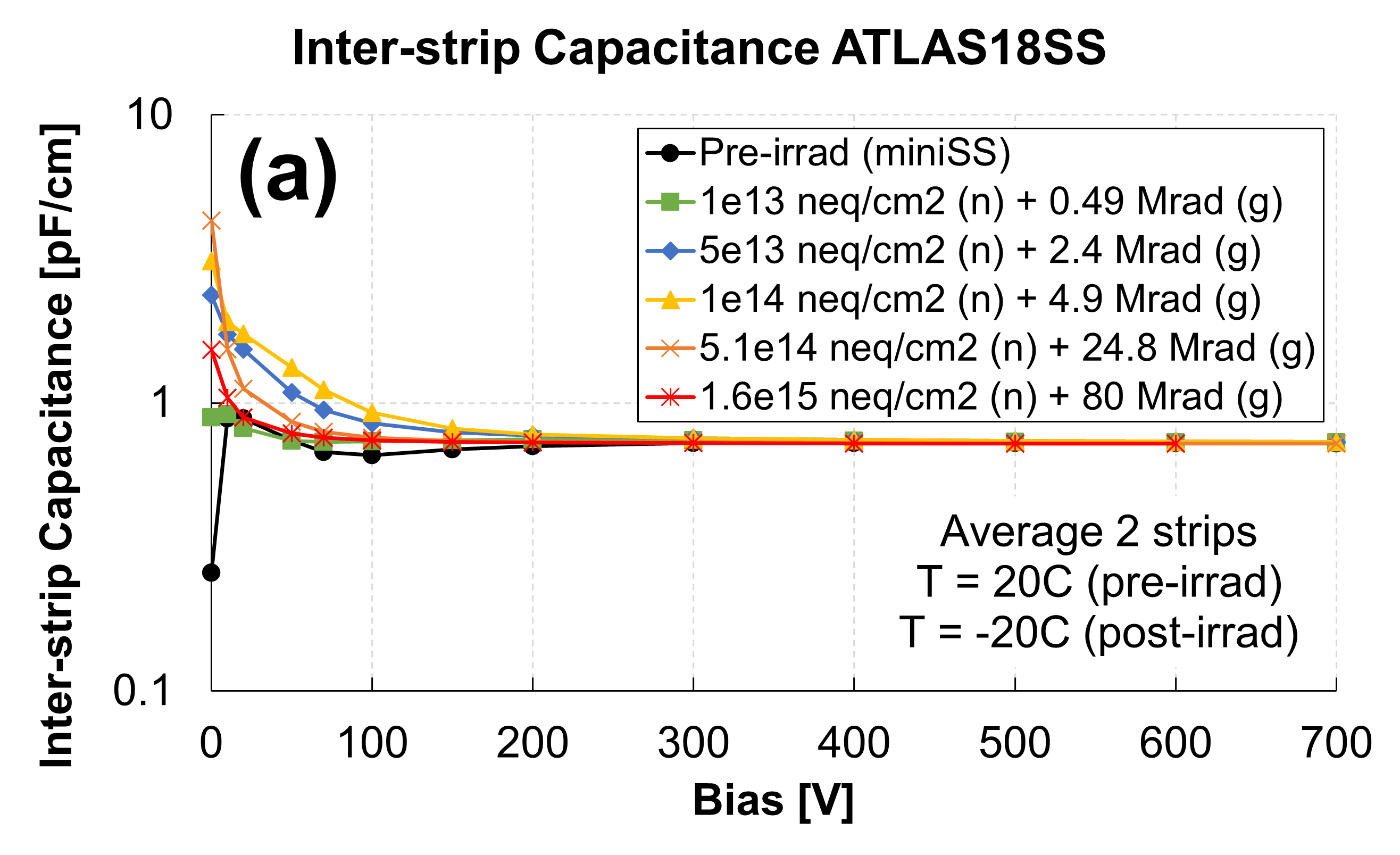}
  \includegraphics[width=0.4\textwidth]{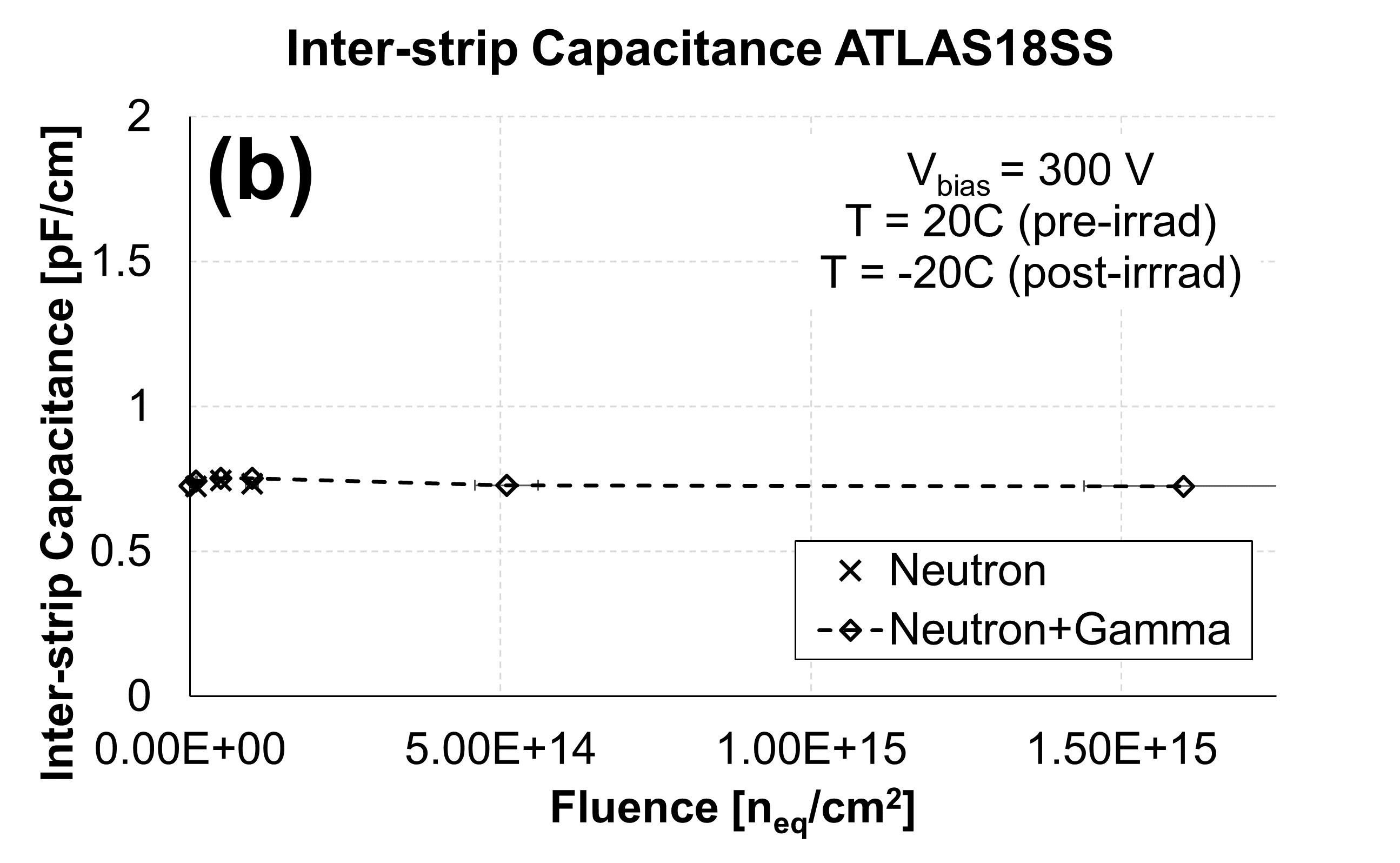}
{\caption{Inter-strip capacitance of neutron+gamma irradiated ATLAS18SS sensors as a function of bias voltage (a) and fluence (b). \label{fig:Chapter2:cint}}}
\end{figure}

For the characterization of the inter-strip capacitance, 2 strips were measured for each fluence and bias voltage, except for -300 V for which 30 strips across each sensor (10 for miniSS sensor) were measured to increase the statistics. Although the inter-strip capacitance show some variation below -300 V (Figure~\ref{fig:Chapter2:cint}(a)), the values are stable for higher bias voltages, and similar for all the neutron and neutron+gamma irradiations (Figure~\ref{fig:Chapter2:cint}(b)), remaining below 1 pF/cm and fulfilling the specifications. Additionally, no clear variation was observed when comparing sensors irradiated only with neutrons or with neutrons+gammas (Figure~\ref{fig:Chapter2:cint}(b)).

\begin{figure}
  \centering
  \includegraphics[width=0.4\textwidth]{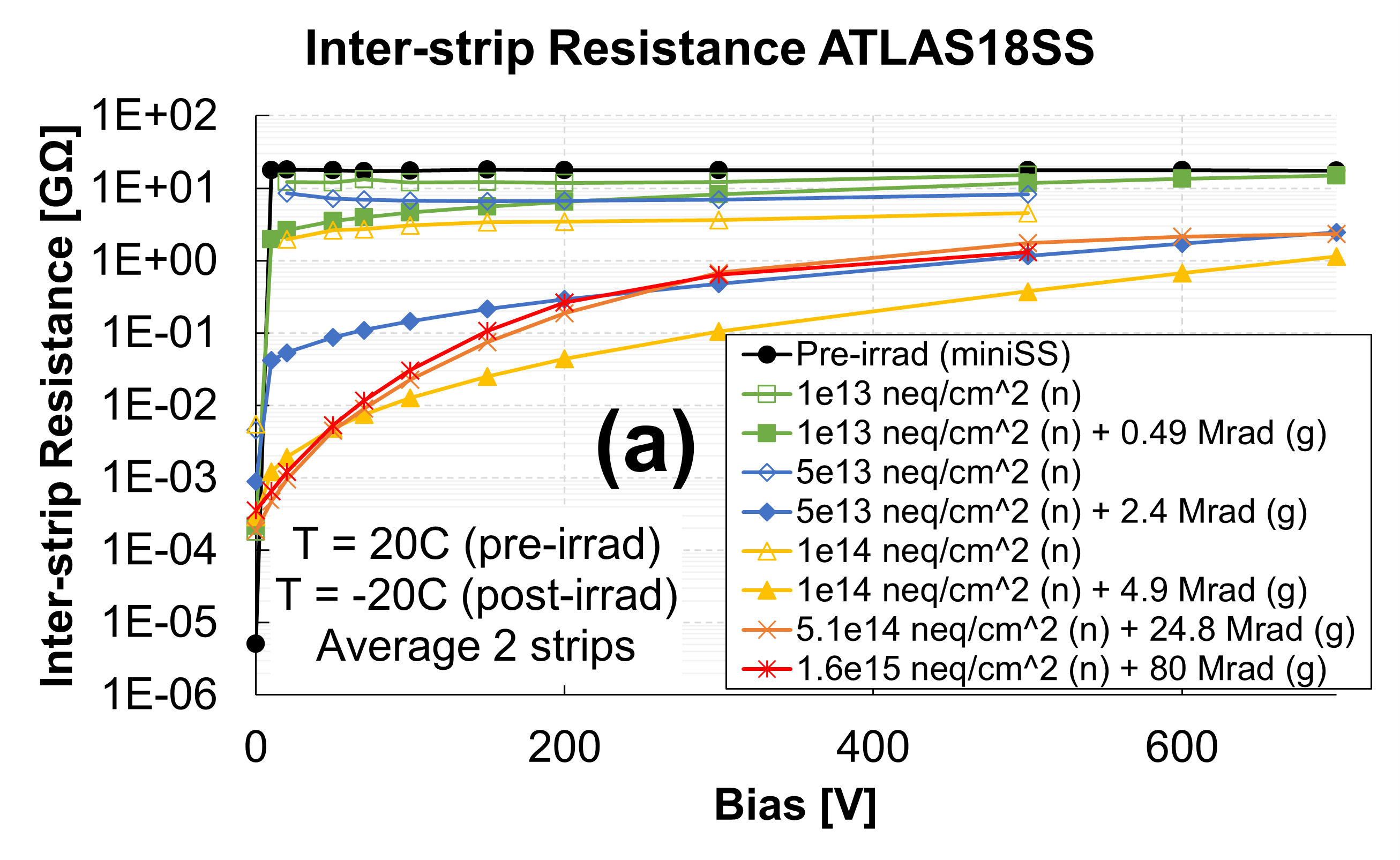}
  \includegraphics[width=0.4\textwidth]{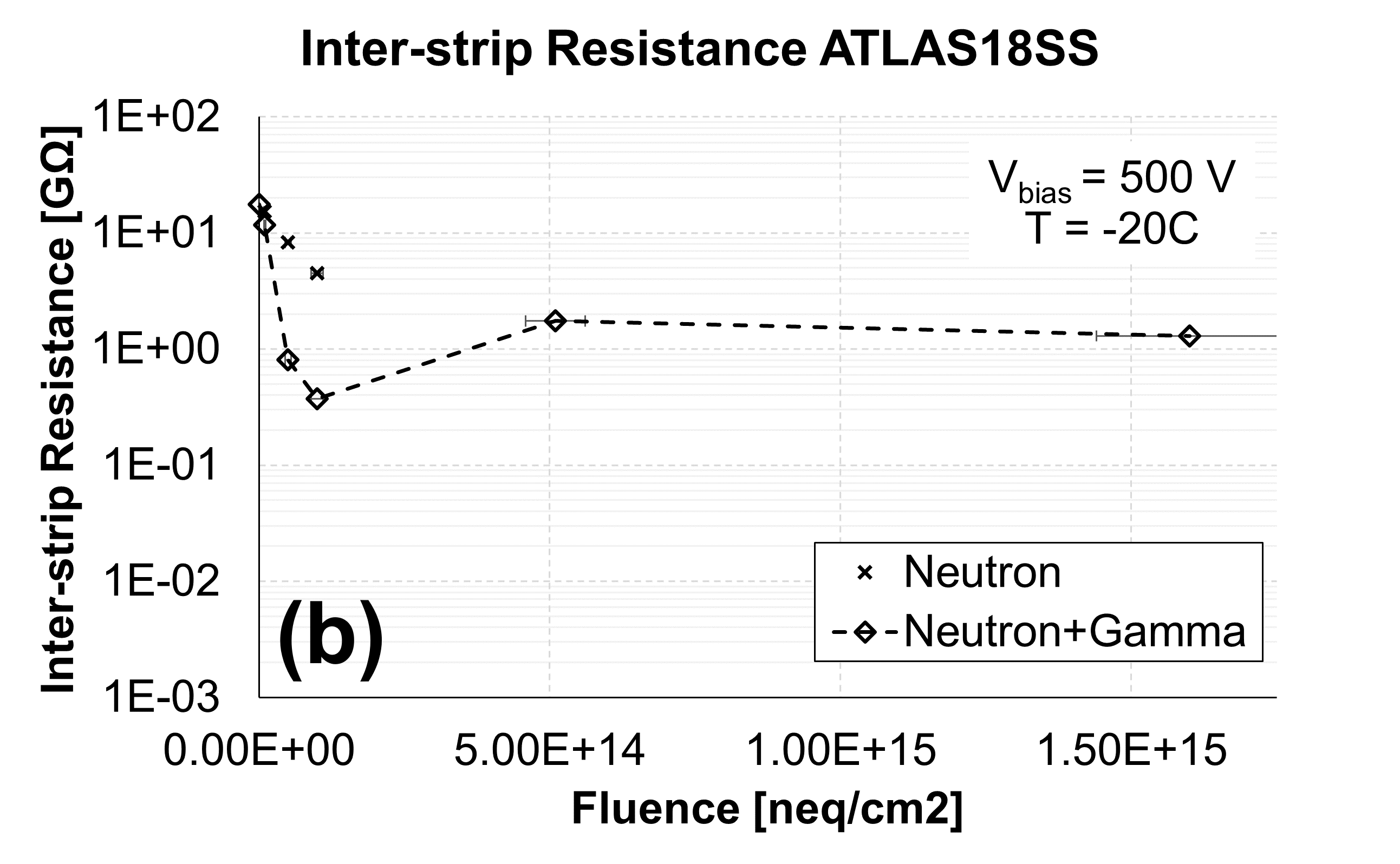}
{\caption{Inter-strip resistance of neutron, and neutron+gamma, irradiated ATLAS18SS sensors as a function of bias voltage (a) and fluence (b). \label{fig:Chapter2:rint}}}
\end{figure}

Similarly, 2 strips per fluence and bias were measured for the characterization of the inter-strip resistance, except for -500 V for which 15 strips across each sensor (10 for miniSS sensor) were measured. The results show a clear reduction of several orders of magnitude as a function of fluence (Figure~\ref{fig:Chapter2:rint}), more pronounced for lower bias voltages, but still within the ATLAS specifications (>10xR$_{bias}$ at -400 V before irradiation, and at -500 V at target fluence), which is consistent with the observations using ATLAS18 Quality Assurance (QA) test structures irradiated with protons~\cite{EricQA}. On the other hand, fluences below 5x10$^{14}$ n$_{eq}$/cm$^{2}$, not studied previously with neutron-irradiated prototypes, show the lowest values of inter-strip resistance when the neutron-irradiated sensors are also irradiated with gammas (Figure~\ref{fig:Chapter2:rint}(b)). This behaviour could be attributed to the predominant ionizing damage originated by the gamma irradiation, with a clear impact on the strip isolation structures (p-stop). This should be investigated in detail in a future work.

\subsection{Single strip characterization}

\begin{figure}
  \centering
  \includegraphics[width=0.4\textwidth]{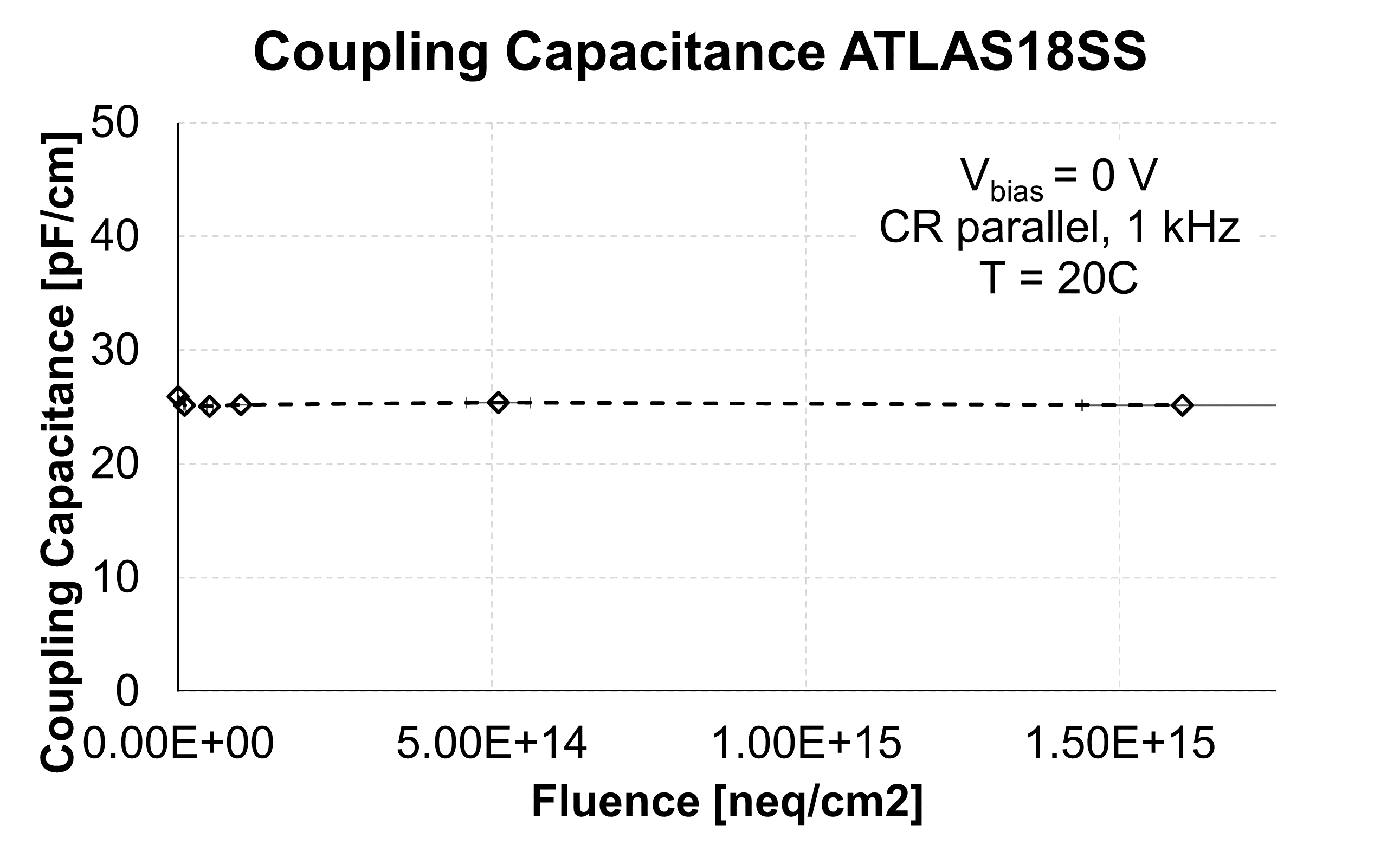}
{\caption{Coupling capacitance of neutron+gamma irradiated ATLAS18SS sensors as a function of fluence. \label{fig:Chapter2:ccoupl}}}
\end{figure}

Similarly to the inter-strip characterization, the coupling capacitance was measured for a total of 30 strips (5 for miniSS sensor) across the sensor for each fluence to increase the statistics. In this case, the measurements were done at room temperature before and after irradiation, as the sensors were tested unbiased. The results obtained (Figure~\ref{fig:Chapter2:ccoupl}) show very stable coupling capacitance values, even for the highest fluences, and above the 20 pF/cm threshold established by the specifications.

\begin{figure}
  \centering
  \includegraphics[width=0.4\textwidth]{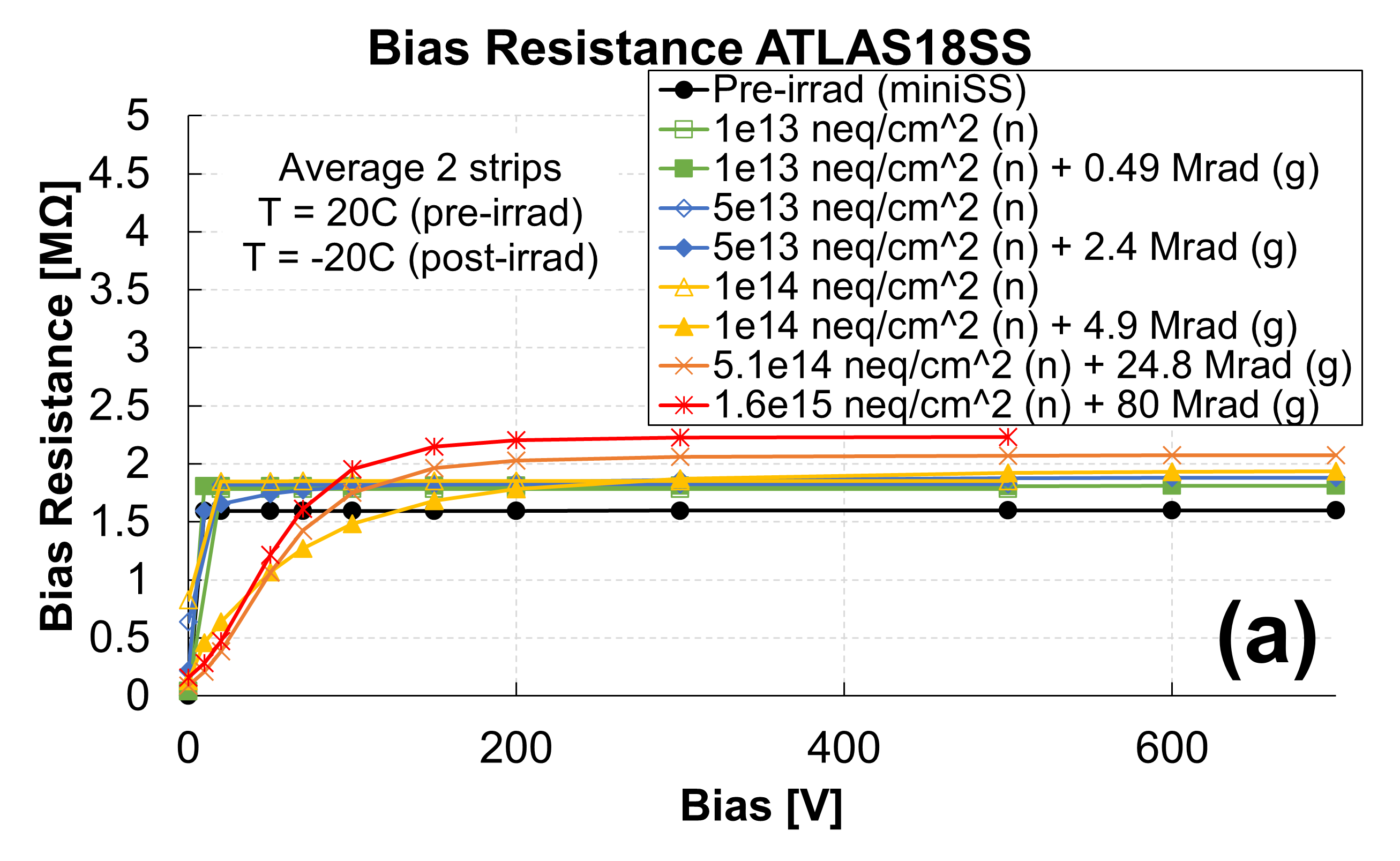}
  \includegraphics[width=0.4\textwidth]{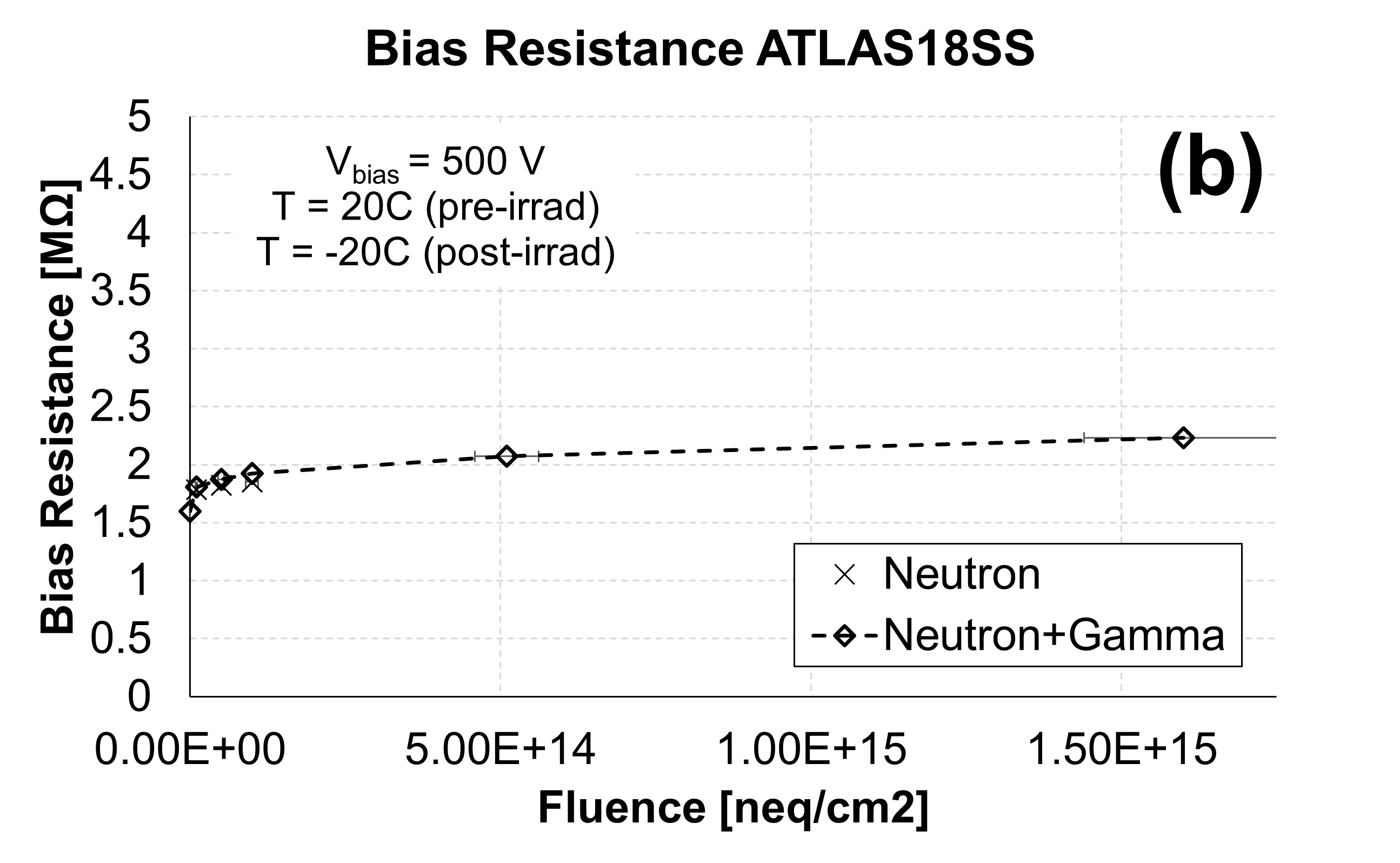}
{\caption{Bias resistance of neutron+gamma irradiated ATLAS18SS sensors as a function of bias voltage (a) and fluence (b). \label{fig:Chapter2:rbias}}}
\end{figure}

As detailed in Section~\ref{sec:tests}, the bias resistance was measured with the same set-up used for the inter-strip resistance, but in this case measuring the current in the strip under test. 2 strips were measured per fluence and bias voltage, except for -500 V for which 15 strips were tested across each sensor (10 for miniSS sensor). The bias resistance values were temperature corrected~\cite{rbias-temperature}. They show some increase with irradiation (Figure~\ref{fig:Chapter2:rbias}), but still fulfilling the ATLAS specifications at the target fluence (1.8$\pm$0.5 M$\Omega$), as previously observed with proton irradiated QA test structures~\cite{EricQA}. An interesting observation is the lower bias resistance values observed at low bias voltages for neutron-irradiated sensors also irradiated with gammas. This behaviour can be attributed to an artifact of the measurement, due to the reduction of the inter-strip resistance caused by the ionizing damage of the gamma irradiation, as discussed in the previous section (Figure~\ref{fig:Chapter2:rint}(b)).

\begin{figure}
  \centering
  \includegraphics[width=0.3\textwidth]{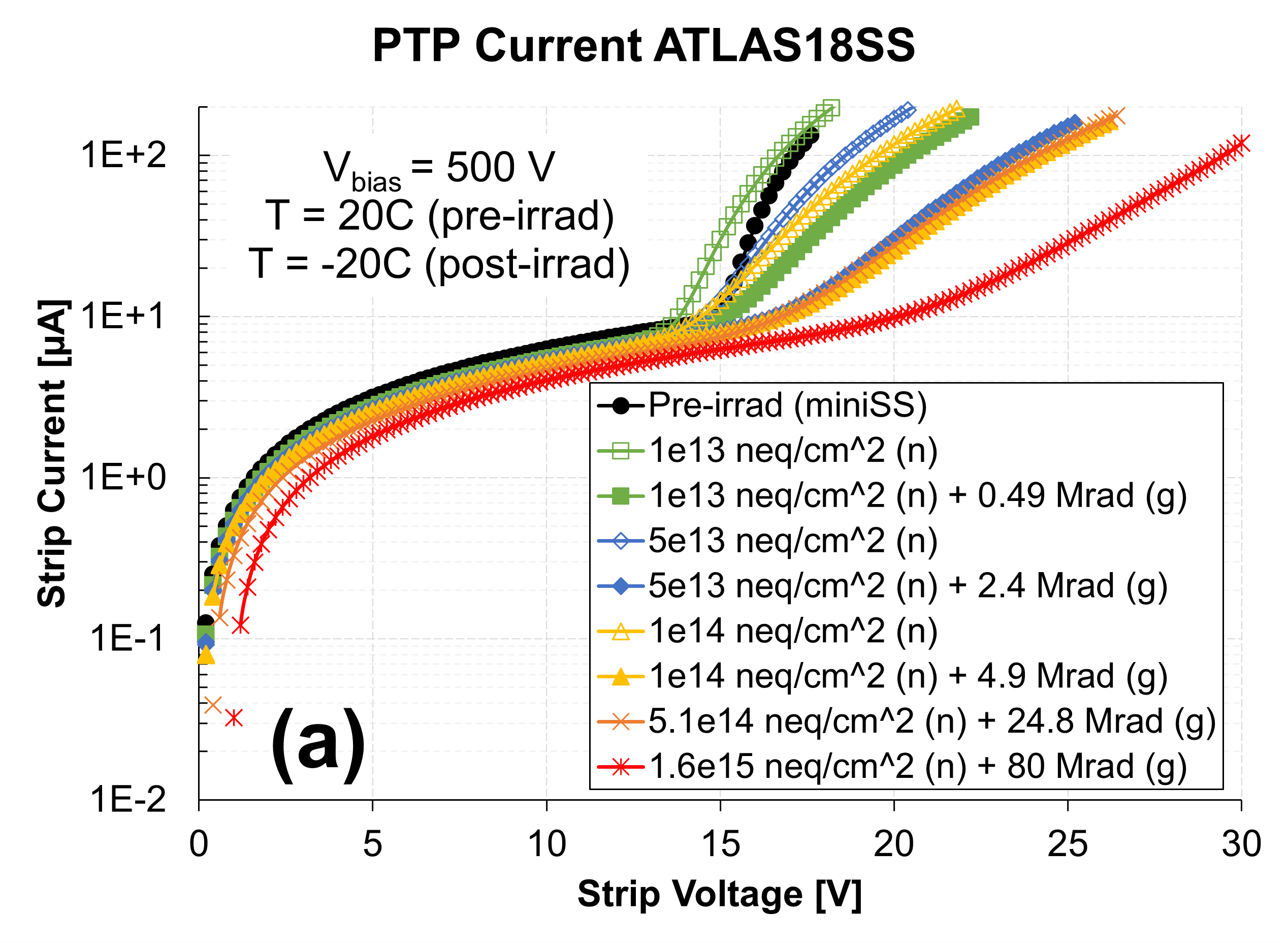}
  \includegraphics[width=0.3\textwidth]{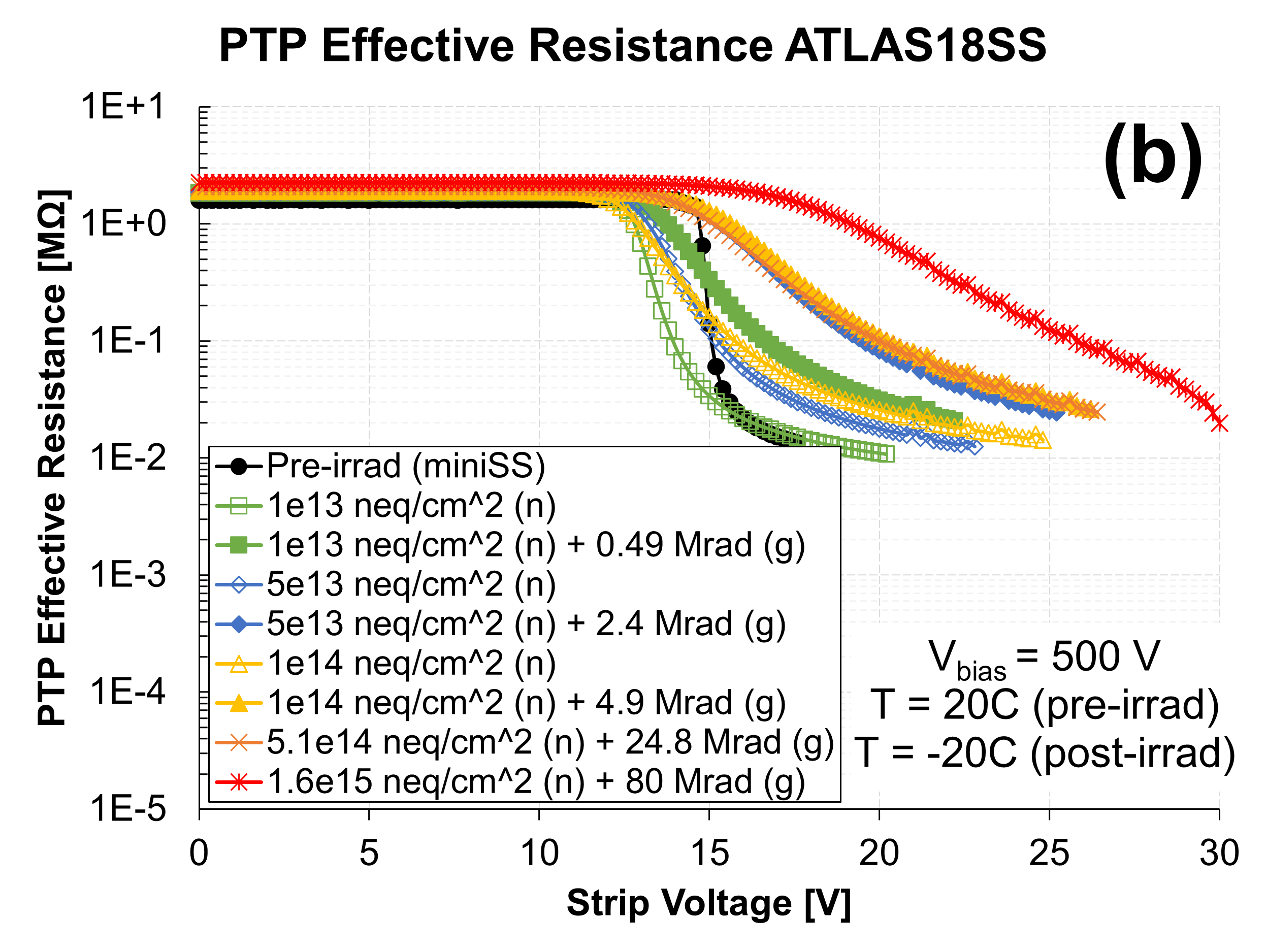}
  \includegraphics[width=0.35\textwidth]{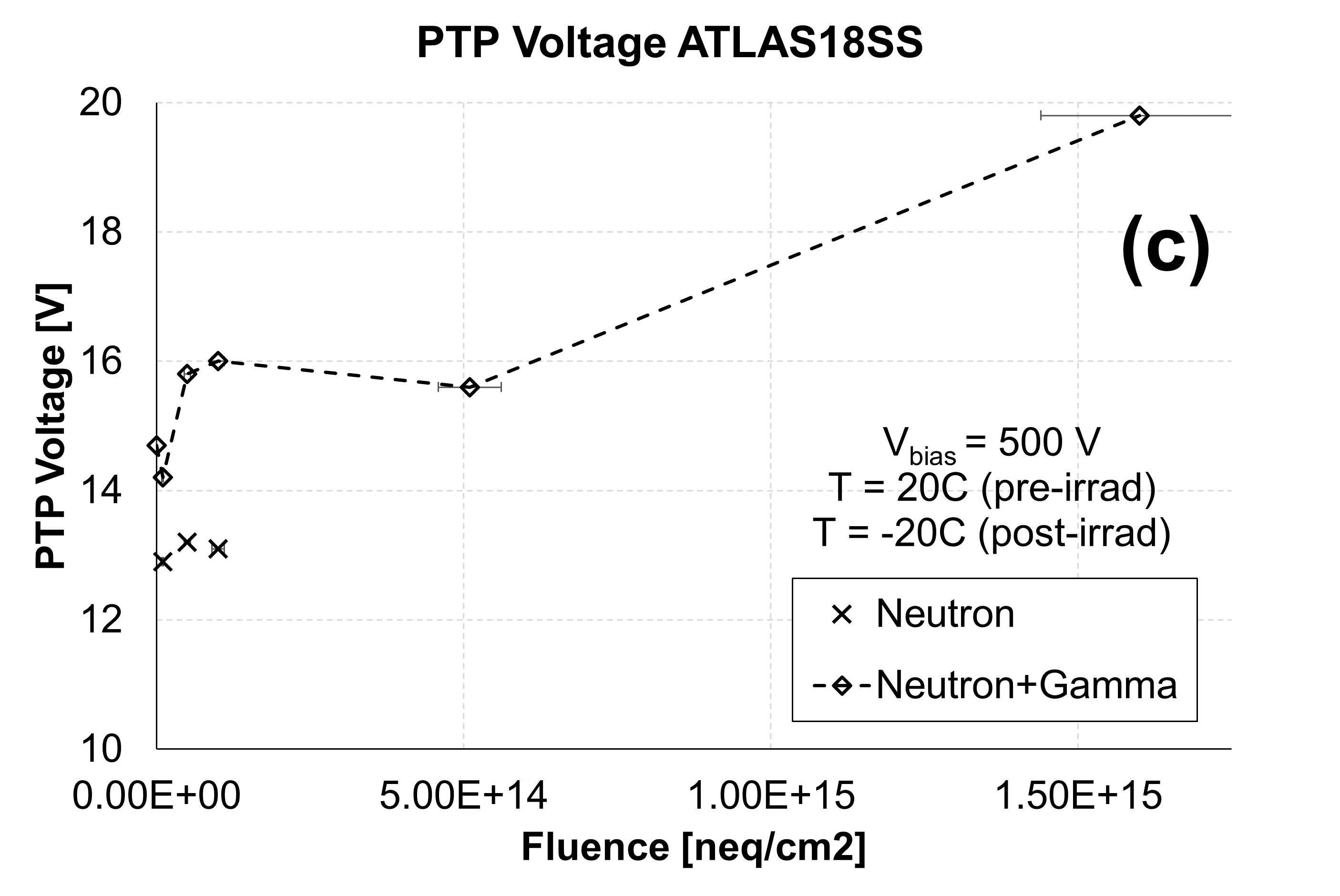}
{\caption{Strip current (a), PTP effective resistance (b) and PTP voltage as a function of fluence (c) of neutron, and neutron+gamma, irradiated ATLAS18SS sensors. \label{fig:Chapter2:ptp}}}
\end{figure}

Finally, the evolution of the punch-through protection, as a function of irradiation, was evaluated measuring 10 strips per fluence, with the sensors biased at -500 V. An increase of the punch-through (PT) voltage with neutron+gamma irradiation is observed (Figure~\ref{fig:Chapter2:ptp}), also seen with QA test structures irradiated with protons~\cite{EricQA}, but showing some decrease at 5x10$^{14}$ n$_{eq}$/cm$^{2}$. This behaviour could be attributed to the transition of the dominant radiation effect on sensors irradiated with neutrons+gammas, from displacement to ionizing damage. Interestingly, sensors irradiated only with neutrons show lower PT voltage values than sensors also irradiated with gammas (Figure~\ref{fig:Chapter2:ptp}(c)), suggesting that ionizing damage increases the PT voltage, although further investigations are needed to understand this behaviour of the irradiated PTP structure.

\section{Conclusions}

This study evaluates the evolution of the electrical characteristics of the new ATLAS ITk strip sensors (ATLAS18) for a wide range of fluences and doses, from the ones expected in the first days of operation (1x10$^{13}$ n$_{eq}$/cm$^{2}$) to the end of the lifetime of the HL-LHC (1.6x10$^{15}$ n$_{eq}$/cm$^{2}$), including a 1.5 safety factor. 

The ATLAS18SS sensors used for this study show a linear increase of the leakage current as a function of fluence, consistent with the current related damage rate expected, and an increase of the full depletion voltage. Fluences below 5x10$^{13}$ n$_{eq}$/cm$^{2}$ show some reduction of the voltage needed to fully deplete the sensor, which could be attributed to the "acceptor removal" effect. A detailed study of the bulk capacitance at different frequencies also confirms that low frequencies ($<$1 kHz) are needed to interact with the deep traps created in highly irradiated sensors, as suggested in previous studies with lower fluences, and confirmed here for the wide range expected for the HL-LHC. The expected under-depletion has known consequences for the signal formation. However, the readout noise in the tracker is affected by the high-frequency capacitance, which does not change with fluence for heavily irradiated sensors. The inter-strip resistance shows an important decrease of several orders of magnitude as a function of fluence, with a clear influence of ionizing damage for lower doses, while the inter-strip capacitance remains low and stable for the range of fluences studied. On the other hand, the bias resistance shows some increase with irradiation, and the coupling capacitance has no clear influence from neutrons and gammas even for the highest fluences. Finally, an increase of the punch-through voltage as a function of neutron+gamma irradiaton was observed, but showing lower values on sensors irradiated only with neutrons, suggesting that the ionizing damage coming from gamma irradiations increases the punch-through voltage.

The results obtained here with full-size ITk strip sensors with final layout design, complement and confirm the studies performed during the development phase with prototype sensors~\cite{MarcelaA17}. All the electrical parameters measured fulfil the specifications established by the ATLAS collaboration~\cite{UnnoPre-production}. Additionally, the neutron+gamma split irradiation, and the study of sensors irradiated to low fluences and doses ($<$5.1x10$^{14}$ n$_{eq}$/cm$^{2}$ and 24.8 Mrad), provide new results about the frequency dependence of the bulk capacitance, and evolution of the inter-strip resistance and punch-through protection, which require further investigation.

\acknowledgments

The work at SFU and TRIUMF was supported by the Canada Foundation for Innovation and the Natural Science and Engineering Research Council of Canada. The work at IMB-CNM is part of the Spanish R\&D grant PID2019-126327OB-C22, funded by MCIN/AEI/10.13039/501100011033 and by ERDF/EU. The work at SCIPP was supported by the US Department of Energy, grant DE-SC0010107. The work at Prague was supported by the European Structural and Investment Funds and the Czech Ministry of Education, Youth and Sports of the Czech Republic via projects LM2023040 CERN-CZ, LTT17018 Inter-Excellence, and FORTE - CZ.02.01.01/00/22-008/0004632. The authors would like to thank the crew at the TRIGA reactor in Ljubljana for help with irradiations. The authors acknowledge the financial support from the Slovenian Research and Innovation Agency (research core funding No. P1-0135 and project No. J1-3032). The authors would like to thank the technical team at the Cyclotron and Radioisotope Center (CYRIC) of Tohoku University (Japan) for proton irradiation. This work was supported by JSPS KAKENHI Grant Number 23K13114.




\begin{thebibliography}{99}

\bibitem{UnnoPre-production}
Y. Unno, et al., \emph{JINST} {\bf 18} (2023) T03008
doi: 10.1088/1748-0221/18/03/T03008


\bibitem{HLLHC}
L. Rossi, et al., \emph{CERN} CERN-ATS-2012-236 (2012)
url: https://cds.cern.ch/record/1471000


\bibitem{ATLASTDR-strips}
ATLAS Collaboration, \emph{CERN} CERN-LHCC-2017-005 (2017)
url: https://cds.cern.ch/record/2257755






\bibitem{XaviThesis}
J. Fernandez-Tejero, \emph{PhD Thesis - UAB}, \emph{JINST} {\bf TH 004} (2020)
url: https://cds.cern.ch/record/2722118


\bibitem{XaviIFX}
J. Fernandez-Tejero, et al., \emph{NIM A} {\bf 981} (2020) 164536
doi: 10.1016/j.nima.2020.164536


\bibitem{MarcelaA17}
M. Mikestikova, et al., \emph{NIM A} {\bf 983} (2020) 164456
doi: 10.1016/j.nima.2020.164456


\bibitem{PTP}
H.F.-W. Sadrozinski, et al., \emph{NIM A} {\bf 699} (2013) 31-35
doi: 10.1016/j.nima.2012.04.062


\bibitem{damage-rate}
M. Moll, et al., \emph{NIM A} {\bf 426} (1999) 87-93
doi: 10.1016/S0168-9002(98)01475-2


\bibitem{frequencyCV}
D. Campbell, et al., \emph{NIM A} {\bf 492} (2002) 402-410
doi: 10.1016/S0168-9002(02)01353-0



\bibitem{acceptor-removal}
S. Terada, et al., \emph{NIM A} {\bf 383} (1996) 159-165
doi: 10.1016/S0168-9002(96)00748-6


\bibitem{frontend-frequency}
K.J.R. Cormier, et al., \emph{JINST} {\bf 16} (2021) P07061
doi: 10.1088/1748-0221/16/07/P07061


\bibitem{EricQA}
E. Bach, et al., \emph{NIM A} {\bf 1064} (2024) 169435
doi: 10.1016/j.nima.2024.169435


\bibitem{rbias-temperature}
V. Latonova, et al., \emph{NIM A} {\bf 1050} (2023) 168119
doi: 10.1016/j.nima.2023.168119






\end{thebibliography}
\end{document}